# Interaction of magnetic-dipolar modes with microwave-cavity electromagnetic fields

E.O. Kamenetskii [1] [*], A.K. Saha [2], and I. Awai [3]

[1] Department of Electrical and Computer Engineering,
Ben Gurion University of the Negev, Beer Sheva, 84105, Israel

[2] Celestica Inc., 844 Don Mills Road, Toronto, ON, M3C 1V7, Canada.

[3] Faculty of Science and Technology,
Ryukoku University Seta, Otsu 520-2194, Japan

We discuss the problem of magnetic-dipolar oscillations combined with microwave resonators. The energy density of magnetic-dipolar or magnetostatic (MS) oscillations in ferrite resonators is not the electromagnetic-wave density of the energy and not the exchange energy density as well. This fact reveals very special behaviors of the geometrical effects. Compared to other geometries, thin-film ferrite disk resonators exhibit very unique interactions of MS oscillations with the cavity electromagnetic fields. MS modes in a flat ferrite disk are characterized by a complete discrete spectrum of energy levels. The staircase demagnetization energy in thin-film ferrite disks may appear as noticeable resonant absorption of electromagnetic radiation. Our experiments show how the environment may cause decoherence for magnetic oscillations. Another noticeable fact is experimental evidence for eigen-electric-moment oscillations in a ferrite disk resonator.

[*] e-mail address: kmntsk@ee.bgu.ac.il

*Introduction.* – In quasi-two-dimensional systems, the dipolar interaction can play an essential role in determine the magnetic properties. In these systems the short-range exchange interactions alone are not necessarily sufficient to establish a ferromagnetically ordered ground state. The dipolar interaction is important in stabilizing long-range magnetic order in two-dimensional systems, as well in determining the nature and morphology of the ordered states. The long-range nature of the dipolar interaction manifests itself in the low-energy excitations of the system [1]. It was shown recently that in a thin-film ferrite disk one can formulate the energy eigenvalue problem for the MS-potential wave functions describing magnetic dipolar modes [2]. For such complex scalar functions the wave equation is the Schrödinger-like equation. This fact allows analyzing the MS oscillations similarly to quantum mechanical problems and gives a basis for a clearer understanding the nature of the observed multi-resonance spectrum. As a feature of such oscillations there are also specific surface magnetic currents, which cause the parity-violating perturbations. As it was theoretically predicted in [3], such processes should lead to appearance of the eigen electric moments (the anapole moments) in disk-form MS-wave ferrite particles. In this paper we give experimental evidence for special interactions between magnetic-dipolar modes oscillating in a ferrite disk and electromagnetic fields of a microwave cavity. There are the observations of the line broadening for magnetic-dipolar-mode spectra and the eigen-electric-moment oscillations.

*Observation of the line broadening for magnetic-dipolar-mode spectra.* – As a starting point in our consideration, we analyze an experimental absorption spectrum of a normally magnetized ferrite disk placed in a region of a maximum tangential RF magnetic field in a rectangular-waveguide cavity resonant in the $TE_{101}$ mode at 4.02 GHz. Such a rectangular-waveguide cavity with an indication of the disk position in a maximum RF magnetic field was shown in [4]. Fig. 1 demonstrates an absorption spectrum obtained for the YIG-film ferrite disk (saturation magnetization $4\pi M_0$ = 1780 Gauss, linewidth $\Delta H$ = 0.6 Oe, film thickness = 0.1 mm, diameter = 3 mm, GGG substrate). The experimental spectrum shown in Fig 1 is similar to the well-known experimental spectra given in [5, 6]. The character of the multi-resonance spectra (obtained with respect to a DC bias magnetic field) leads to a clear conclusion that the energy of a source of a DC magnetic field is absorbing "by portions" or discretely, in other words. There are the multi-resonance regular experimental spectra with properties similar to an atomic-like $\delta$ - function density of states. One can discern a striking difference in the picture of absorption spectra for planar ferrite disk and ferrite spheres. In the last case there are only a very few (and rather broad) absorption peaks in a spectrum [7]. Based on the spectrum in Fig. 1, we can estimate the bias-field energies necessary for transitions from the main level to upper levels. For example, to have a transition from the first level to the second level we need the energy surplus:

$$\Delta U_{12} = 4\pi M_0 \left( H_0^{(2)} - H_0^{(1)} \right) = 1780 \times (2920 - 3015) = -1.69 \times 10^4 \frac{Joule}{m^3}$$

The breadth of a spectral line for atoms and molecules is determined by broadening of energy levels due to the external-factor interactions. In our case this breadth should be due to intrinsic relaxation processes in a ferrite material and the radiation damping of a whole ferrite sample. Based on a classical electromagnetic theory of a precessing point magnetic dipole, it was shown in [8] that magnetic-dipolar radiation damping can be a primary source of the broadening of the *uniform mode* in ferromagnetic resonance. One can suppose that the pronounced *first-peak* line broadening, which we observed in [4] for ferrite disks with larger diameters, is due to the magnetic-dipolar radiation damping, causing a cavity overloading at the resonance frequency (or the resonance DC magnetic field) of the main oscillating mode. This mechanism cannot give, however, any explanations for possible line broadening of *high-order* peaks in a spectrum.

External interactions increasing the probability of transition of a quantum system to other states lead to line broadening. To enhance this interaction in our case we followed the ideas that (a) the electromagnetic field should be more intensified in a region of a ferrite particle and (b) a magnetic particle should be properly matched with a cavity. To satisfy these conditions we used a double-ridged-waveguide cavity with tapered fins. This cavity is shown in Fig. 2. At the end opposite to a short wall we put an iris (not shown in Fig. 2). The bias magnetic field $H_0$ and the disk axis were oriented along *y*-axis. In experiments we used the $TE_{102}$ cavity mode with resonance frequency 3.735 GHz. For a ferrite disk, having 3mm diameter, placed in position #1 (see Fig. 2) corresponding to a maximum RF tangential magnetic field $H_x$, we obtained the spectrum shown in Fig. 3. It is interesting also to consider analogous spectrum for a disk with another diameter. Fig. 4a shows the spectrum obtained in a maximum RF magnetic field (position # 1 in Fig. 2) for a disk having 5mm diameter. Fig. 4b demonstrates the high-order-peak part of the same spectrum in an extended scale of the bias magnetic field.

One can see that a character of spectrums observed in Figs. 3 and 4a, b is strongly different from spectrums shown in Fig. 1. There are no clearly distinguishable "levels of reference" for resonance peaks. Compared to the atomic-like ($\delta$-functional) character of a spectrum in Fig. 1, there is the *pronounced line broadening for all of the peaks in the multi-resonance spectrum*. Moreover, the broadening of energy levels can be compared with the difference of energy between levels.

As a ferrite specimen changes its position along z-axis in a cavity with tapered fins, the tangential RF magnetic field $H_x$ decreases. Figs. 5a and 5b show spectrums obtained for the 3mm-diameter disk placed in a cavity, respectively, in positions ## 2 and 3 (see Fig. 2). It is evident that as a specimen moves from position # 1, the radiation damping decreases and so the line broadening decreases as well. We restore the line spectra. Comparing the spectrum in Fig. 3 with the spectra in Figs. 5 a, b, one can see that instead of asymmetric positions of neighboring peaks in Fig. 3, the same peaks in Figs. 5 a, b are symmetrically oriented. It is necessary to note also that compared to the upward-peak absorption spectrum for a rectangular-waveguide cavity shown in Fig. 1, for double-ridged-waveguide cavity we have the downward-peak absorption spectrum.

*Obsevation of the eigen-electric-moment oscillations in ferrite disk resonators.* – The theoretical analysis [3] predicts that the eigen-electric-moment oscillations should be observed in a normally magnetized small ferrite disk. So one can expect that ferrite resonator will be strongly affected by the external normal RF electric field with a clear evidence for *multi-resonance* oscillations. It can be easily proved that the eigen electric moment of a ferrite disk arises not from the classical curl electric fields of MS oscillations (which give zero electric moments of a sample). At the same time, any *induced* electric polarization effects in YIG or GGG materials are beyond the frames of the spectral properties of MS oscillations.

To get a direct experimental proof of the eigen-electric-moment oscillations in ferrite disk resonators we use a careful *comparative analysis* of the experimental spectra obtained for different-type (magnetic, electric, or combined: electric and magnetic) exciting fields. To have necessary clearance in the field structure in a cavity we used a double-ridge-waveguide cavity without tapered fins. Such a cavity allows achieving the enhanced sample-cavity interaction with definite correlation between the disk position and the cavity field. The cavity is shown in Fig. 6a. The calculated field structures of the cavity $TE_{102}$ mode are shown in Figs. 6b, c. The cavity resonance frequency is 3.76 GHz.

A normally magnetized disk-form YIG film ferrite resonator (saturation magnetization $4\pi M_0 = 1780$ Gauss, linewidth $\Delta H = 0.6$ Oe, film thickness = 0.1 mm, diameter = 3 mm, GGG substrate) was placed in different positions in a cavity. In the position near a short-circuited wall (position #1 in Fig. 6), MS oscillations are excited by the maximum tangential RF magnetic field $H_x$. In this position the normal RF electric field $E_y$ is equal to zero. Fig. 7 (a) demonstrates the observed spectrum of MS oscillations. As a ferrite specimen changes its position along z-axis from the short-circuited wall to the point quarter wavelength far, the tangential RF magnetic field $H_x$ decreases and the normal RF electric field $E_y$ increases. For different positions ## 2, 3 and 4 in Fig. 6, the corresponding oscillating spectra are shown, respectively, in Figs. 7 (b), 7 (c) and 7 (d). In position # 4 we have the maximum normal electric field $E_y$ and zero tangential magnetic field $H_x$.

A comparative analysis of spectra in Figs. 7 (a), 7 (b), 7 (c) and 7 (d) demonstrates the fact of excitation of multi-resonance oscillations in a ferrite disk resonator by a *normal RF electric field*. Making this comparative analysis one may suppose, however, that the progressive appearance and increase of additional absorption peaks are due to the increasing role of the inhomogeneous tangential RF magnetic field, but not due to the role of the normal electric field. It becomes clear, however, that such the model cannot be accepted. Passing from Fig. 7 (a) to Fig. 7 (d), we do not see a real *tendency of the dominant role* of the inhomogeneous tangential magnetic field. If such a tendency appears, the peaks excited by the homogeneous RF magnetic field [shown in Fig. 7(a)] should become more and more suppressed. Evidently, this situation does not take place.

*Discussions.* – The theory [2] shows that the confinement effect leads to the quantum-like effects for MS oscillations in a ferrite disk. MS oscillations can be considered as the motion process of certain quasiparticles – the light magnons – having quantization of energy and characterizing by effective masses depending on energy levels. To a certain extent an analysis of magnetic-dipolar ferrite disks [2] resembles study of the eigenvalue problem for the Schrödinger operator on a two dimensional semiconductor disk [9]. For a case of a magnetic-dipolar-mode ferrite disk one has the quantized-like oscillating system which preserves the coherence. Absorption spectrum in Fig. 1 shows a series of sharp field-dependent resonances. It means that the ferrite-particle-cavity-field entanglement can be strongly reduced. Following an idea that external interactions should increase the probability of transition of a quantum system to other states leading to line broadening, we showed here how the environment may cause decoherence for magnetic oscillations. By sweeping the bias magnetic field at a certain frequency, we monitor oscillations. Many fringes become visible. A closer look at the lower-field region in Figs. 3, 4 shows regular changes in the oscillation amplitude and reveals the appearance of additional periodic oscillations between the main resonances. Can the observed multiresonance spectra shown in Figs. 3, 4 be just only due to the classical effect of over-coupling in the cavity-ferrite disk system? As an answer to this question it is necessary to point out that the classical over-coupling does not give sharp deep downfalls between resonances, as it takes place in our case.

Together with such similarity with semiconductor quantum dots as discrete energy levels due to confinement phenomena, ferrite particles show other, very unique, properties attributed to the quantized-like systems. There are special symmetry properties of the anapole moments [3]. In this paper we presented experimental results showing that MS oscillations in a normally magnetized ferrite disk are strongly affected by a normal component of the external RF electric field. There are the *eigen-electric-moment* oscillations caused by special motion processes in a ferrite resonator. Since the RF electric field does not change sign under time inversion, the eigen electric moment should also be characterized by the time-reversal-even properties. The character of magnetic oscillations does not exhibit any properties of the *potential* electric fields and, therefore, any multiresonance electric polarization spectrum. The only reason to explain the observed "microwave electric properties" could be found from the circular magnetic currents, which appear due to confinement phenomena of MS oscillations in a ferrite disk [3]. A comparative analysis of spectra in Figs. 7 (a), 7 (b), 7 (c) and 7 (d) demonstrates the fact of excitation of multi-resonance oscillations in a ferrite disk resonator by the normal RF electric field. The observed spectral properties for a ferrite particle placed in a maximum of a homogeneous electric field should, certainly, be due to the electric dipole radiation and not due to the magnetic quadrupole radiation (which one can suppose might appear because of the maximal nonhomogeneity of a cavity magnetic field).

------------------------------------------

**Interaction of magnetic-dipolar modes with microwave-cavity electromagnetic fields**

By E.O. Kamenetskii, A.K. Saha, and I. Awai

**Figure captions**

Fig. 1. Absorption spectrum of a normally magnetized ferrite disk in a rectangular-waveguide cavity (maximum tangential RF magnetic field). In the insertion: a top view of a ferrite disk with orientation of the RF magnetic field.

Fig. 2. Double-ridged-waveguide cavity with tapered fins: (a) rough sketching, (b) top view.

Fig. 3 Absorption spectrum of a normally magnetized 3mm-diameter ferrite disk in a cavity with tapered fins in position # 1 (maximum tangential RF magnetic field).

Fig. 4 (a) Absorption spectrum of a normally magnetized 5mm-diameter ferrite disk in a cavity with tapered fins in position # 1 (maximum tangential RF magnetic field), (b) the high-order-peak part of the same spectrum.

Fig. 5. Absorption spectrum of a normally magnetized 3mm-diameter ferrite disk in a cavity with tapered fins: (a) in position # 2 ( 15 mm far from position # 1), (b) in position # 3 (20 mm far from position # 1).

Fig.6. Double-ridged-waveguide cavity without tapered fins. Experimental arrangement: (a) rough sketching, (b) and (c) the field structure in the cavity.

Fig. 7. Absorption spectrums of a normally magnetized ferrite disk in a double-ridged-waveguide cavity:

(a) in position # 1 (near a short end wall; maximum tangential RF magnetic field),

(b) in position # 2 (17 mm far from a short end wall),

(c) in position # 3 (22 mm far from a short end wall),

(d) in position # 4 (28 mm far from a short end wall; maximum normal RF electric field).

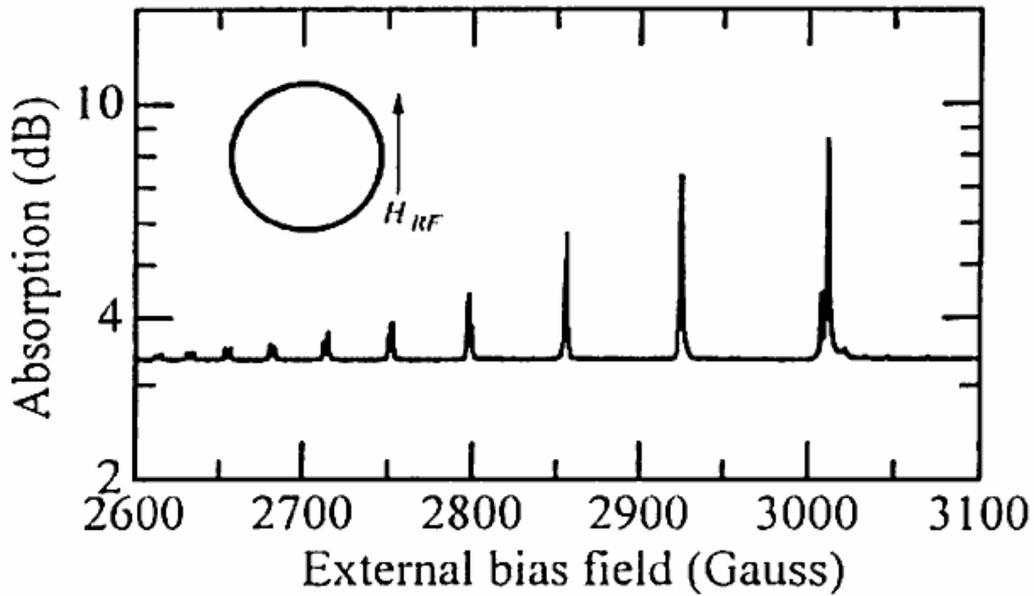

Fig. 1. Absorption spectrum of a normally magnetized ferrite disk in a rectangular-waveguide cavity (maximum tangential RF magnetic field). In the insertion: a top view of a ferrite disk with orientation of the RF magnetic field.

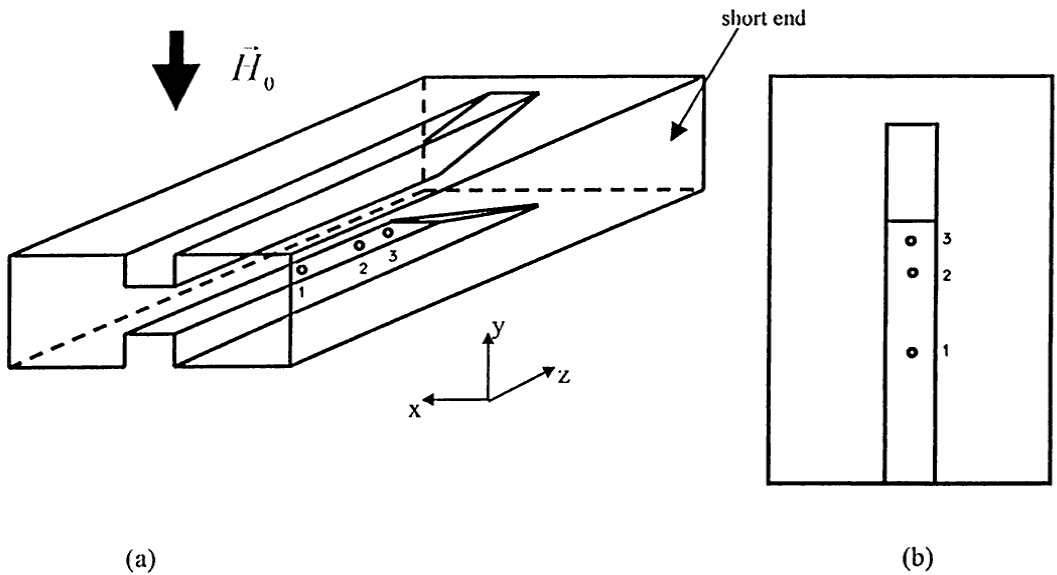

(a)  (b)

Fig. 2. Double-ridged-waveguide cavity with tapered fins: (a) rough sketching, (b) top view.

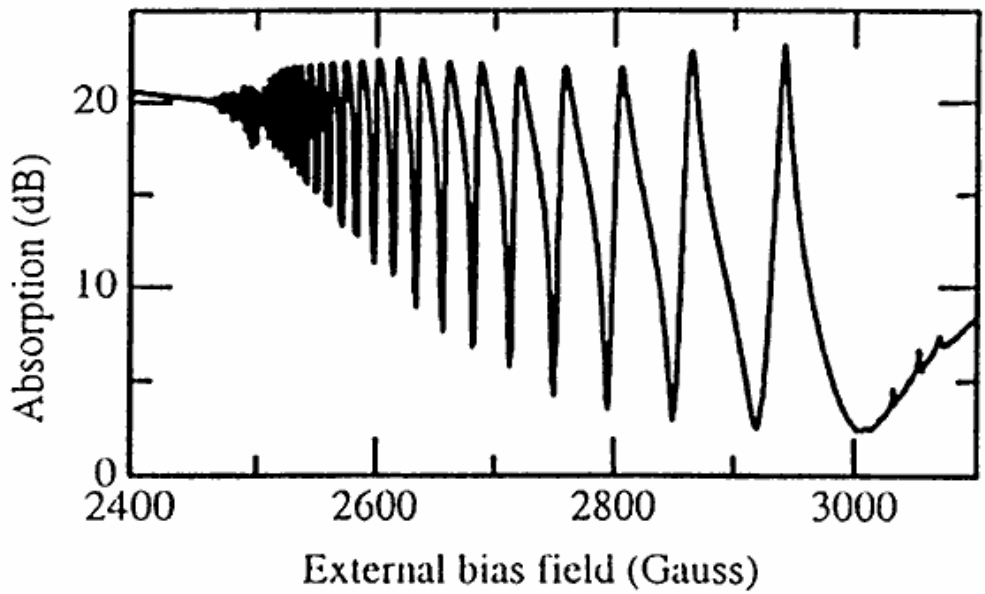

Fig. 3  Absorption spectrum of a normally magnetized 3mm-diameter ferrite disk in a cavity with tapered fins in position # 1 (maximum tangential RF magnetic field).

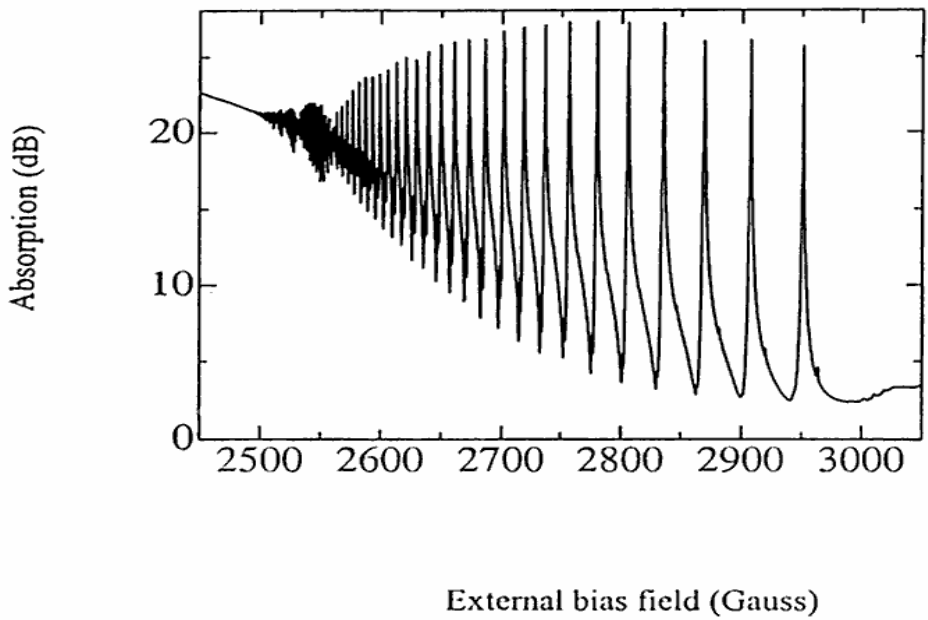

(a)

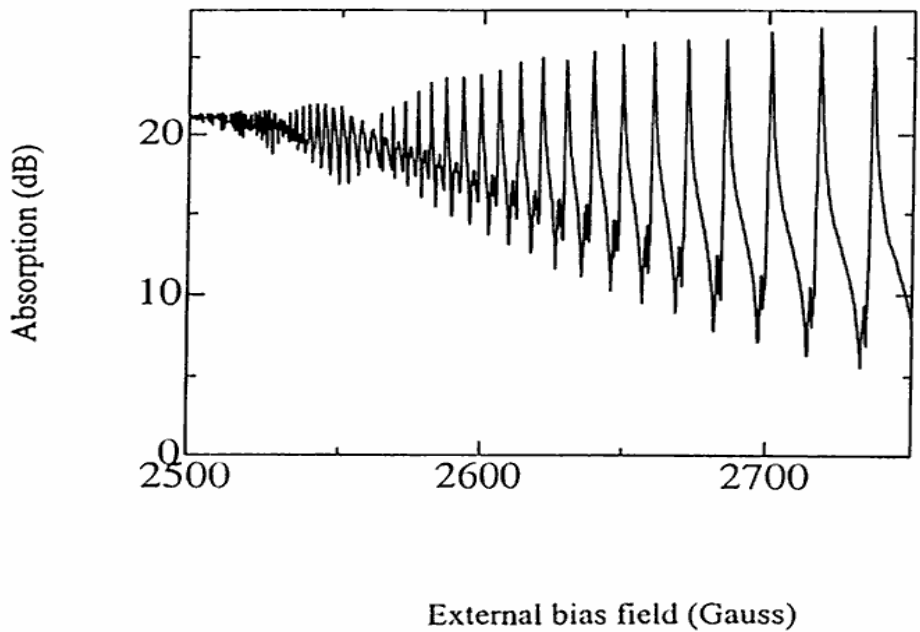

(b)

Fig. 4 (a) Absorption spectrum of a normally magnetized 5mm-diameter ferrite disk in a cavity with tapered fins in position # 1 (maximum tangential RF magnetic field), (b) the high-order-peak part of the same spectrum.

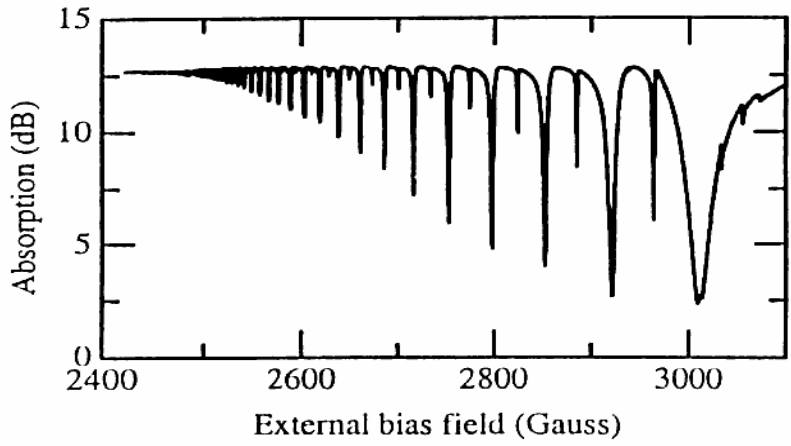

(a)

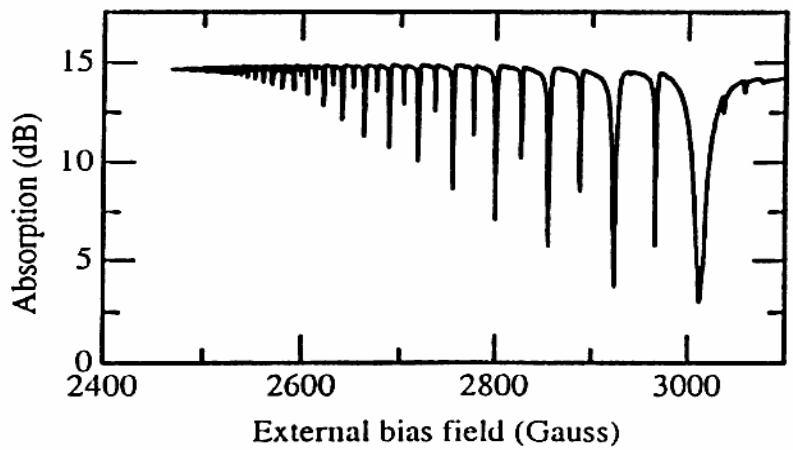

(b)

Fig. 5. Absorption spectrum of a normally magnetized 3mm-diameter ferrite disk in a cavity with tapered fins: (a) in position # 2 ( 15 mm far from position # 1), (b) in position # 3 (20 mm far from position # 1).

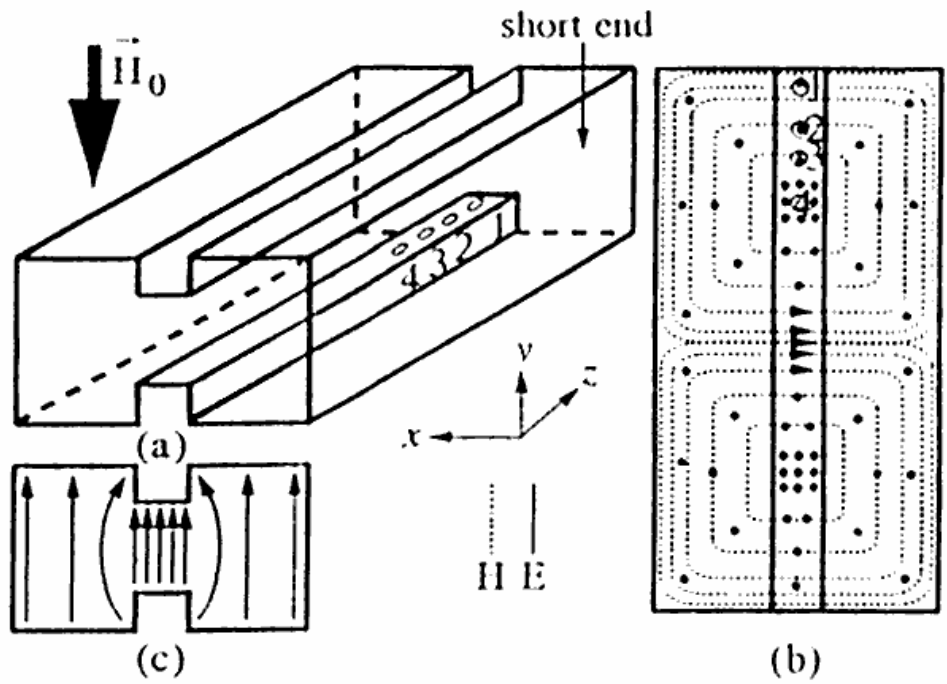

Fig.6. Double-ridged-waveguide cavity without tapered fins. Experimental arrangement: (a) rough sketching, (b) and (c) the field structure in the cavity.

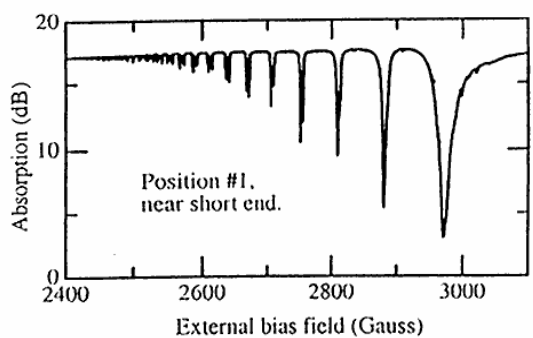
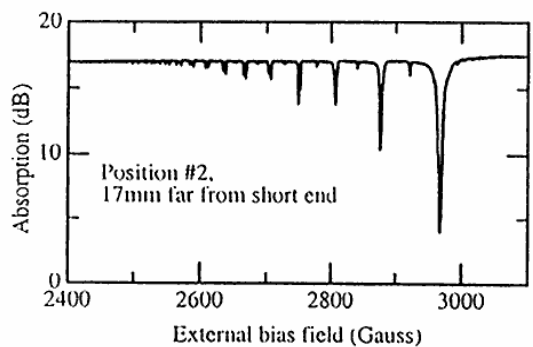

(a)

(b)

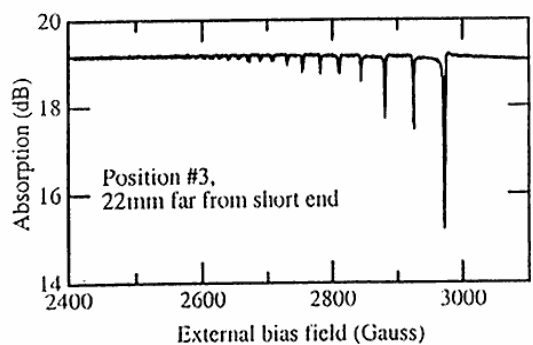
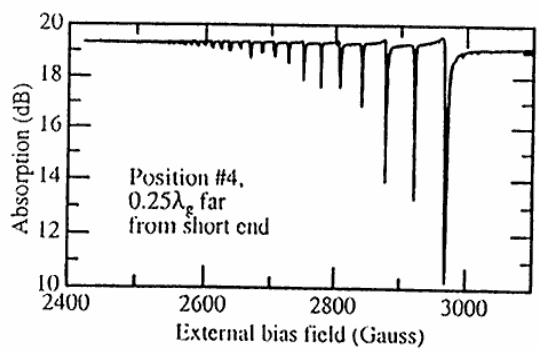

(c)

(d)

Fig. 7. Absorption spectrums of a normally magnetized ferrite disk in a double-ridged-waveguide cavity:

(a) in position # 1 (near a short end wall; maximum tangential RF magnetic field),

(b) in position # 2 (17 mm far from a short end wall),

(c) in position # 3 (22 mm far from a short end wall),

(d) in position # 4 (28 mm far from a short end wall; maximum normal RF electric field).